0

2

# $D-$ AND $B-$MESON SEMI-LEPTONIC DECAYS

ELC (European Lattice Collaboration) and the APE Collaboration
A.Abada[a], C.R.Allton[b], A.Bartoloni[b], C.Battista[b], P.Boucaud[a], S.Cabasino[b], N.Cabibbo[b], D.B.Carpenter[c], M.Crisafulli[b], S.Güsken[d], P.Hernandez[e], V.Lubicz[b], G.Martinelli[b], F.Marzano[b], P.S.Paolucci[b], J.Pech[b], O.Pene[a], F.Rapuano[b], C.T.Sachrajda[f], R.Sarno[b], K.Schilling[d], G.Siegert[d], N.Stella [b], G.M.Todesco[b], M.Torelli[b], W.Tross[b], P.Vicini[b] and A. Vladikas[g].

[a]LPTHE Orsay, France.
[b]Dip. di Fisica, Univ. di Roma, "La Sapienza", Italy.
[c]Dep. of Electronics, The University, Southampton, UK.
[d]Physics Dep., Univ. of Wuppertal, Germany.
[e]Dept de Fís. Teórica, Univ. Autónoma de Madrid, Spain,
[f]Dep. of Physics, The University, Southampton, UK.
[g]Dip. di Fisica, Univ. di Roma, "Tor Vergata", Italy
Talk presented by N. Stella*

Recent results for semi-leptonic decays of $D$ and $B$ mesons at $\beta = 6.4$, using the Wilson action, and at $\beta = 6.0$, using the Clover action, are reported.

Semi-leptonic decays of heavy flavours play a crucial role in the determination of the Cabibbo-Kobayashi-Maskawa (CKM) matrix elements. $D$ decays provide a good test of the method, since the relevant CKM matrix element is well constrained by unitarity in the Standard Model: $V_{cs} \simeq 0.975$. Here we present preliminary results obtained from a large statistics simulation performed by the APE Collaboration, using the Clover action[1], and results obtained by the European Lattice Collaboration (ELC) [2], using the Wilson action at $\beta = 6.4$. The full results of this work will be published elsewhere [3].

Both the studies were intended to reduce $O(a)$ artefacts: the ELC study by working at a smaller value of $a$, but with Wilson fermions; the APE study by working at moderate $a$ but with an "improved" action. The results with the Clover action suffer from severe statistical errors, in spite of the large statistics. There are two possible reasons for this: either the Clover action or the "thinning" procedure. "Thinning" means that we use only one point out of three, in each spatial direction, when we compute correlation functions. This procedure is necessary when, as in our case, the computer memory is not sufficient to store the full quark propagators. There is a systematic error introduced by "thinning" because we cannot eliminate high momentum components in the correlation functions. This error can be shown to be negligible at large time distances. On the other hand, by using a small sample of the points, the signal is destabilized.

Both the studies try to extrapolate the form factors (FF) to $B \to \pi, \rho$ decays using the scaling laws predicted by the Heavy Quark Effective Theory (HQET) [4]. For the moment this is more a feasibility study because of the large statistical errors which amplify in the extrapolation.

Semi-leptonic decays ($D \to K, K^\star$ $B \to D, D^\star$ $B \to \pi, \rho$) are described in terms of six FF of which only four are relevant for the decay rates (see eg. [5]). The relevant information, for each FF, can be expressed in terms of its value at $q^2 = 0$ and its $q^2$ dependence. In the following, we will give the main results of the two studies for these quantities.

*Present address: Dep. of Physics, The University, Southampton, UK

Table 1

| | # of Confs | Action | $\beta$ | Volume $L^3 \times T$ | $a^{-1}$ [GeV] |
|---|---|---|---|---|---|
| ELC [2] | 15 | Wilson | 6.4 | $24^3 \times 60$ | 3.6(2) |
| APE | 100 | Clover | 6.0 | $18^3 \times 64$ | 2.03(9) |

**Computation details**

To extract the FF we have computed two- and three-point correlation functions and followed the procedure set out in [2]. The lattice setup of the two computations is reported in Table 1.

To study the $q^2$ dependence of the FF, we compute the matrix elements with a $D$ meson at rest, i.e. $\vec{p}_D = \vec{0}$, and $\vec{p}_K \equiv 2\pi/La \cdot (0,0,0),(1,0,0),(1,1,0),(1,1,1),(2,0,0)$. In the APE computation, other values of the momentum transfer were studied by fixing $\vec{p}_D \equiv 2\pi/La \cdot (1,0,0)$, and $\vec{p}_K \equiv 2\pi/La \cdot (0,0,0),(1,0,0),(-1,0,0),(0,1,0)$. Correlation functions which are equivalent under the hypercubic symmetry have been averaged together.

**Test of Vector Meson Dominance**

In fig. 1, we compare the FF, computed in the APE simulation, to the meson dominance predictions, e.g. $f^+(q^2) = \frac{f^+(0)}{1-q^2/M^2_{1^-}}$, as a function of $q^2$. In the range explored, the $q^2$ behaviour of the FF is similar to the expectation of meson dominance. From our data, there could be a deviation from meson dominance at $q^2 > 0$, that can be decided only with more accurate results.

To obtain the FF at $q^2 = 0$ it is convenient to estimate them from the points with the $q^2$ closest to zero. In this way one reduces the systematic effect due to the choice of a particular functional dependence on $q^2$.

***D*-meson decays**

We report in Tab. 2 the values of the FF at $q^2 = 0$ obtained by the ELC and the APE Collaborations for the $D \to K(K^*)$ decays. Some results obtained in the past years are also reported for comparison[1]. The numbers reported by ELC are consistent within large uncertainties with previous results obtained at larger values of $a$. The APE results are still preliminary and have been obtained by using two different procedures, the so called "ratio"(A) and "analytic"(B) methods, for details see ref. [2].

[1] Other results presented at this conference [11,12] can be found in these proceedings, see also [13].

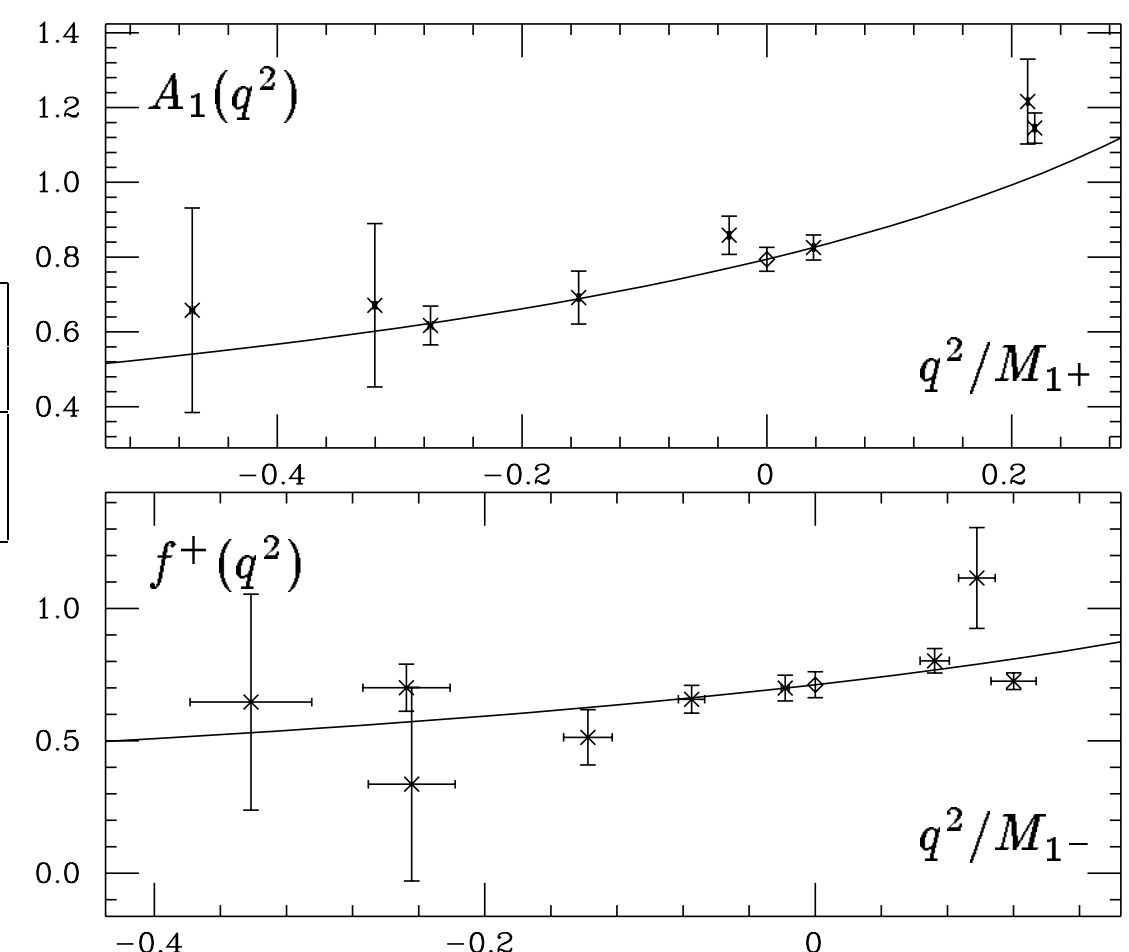

Figure 1. Form Factors $f^+(q^2)$ and $A_1(q^2)$ as a function of the momentum transfer $q^2/M_{\rm pole}$. The curve corresponds to the nearest pole dominance approximation.

One source of uncertainty is the value of the renormalization constants of the vector and axial currents, $Z_V$ and $Z_A$. ELC computed the vector FF by using both the conserved vector current and the local vector current with $Z_V = 0.84$. For the axial current, the perturbative value $Z_A = 0.88$ has been used. Different values should be used if one assumes boosted perturbation theory[14,15]. In the APE simulation we used local currents with $Z_V = 0.824$ and $Z_A = 1.09$, as determined non perturbatively, using light quark correlation functions, see ref. [16,17].

As table 2 shows, there is reasonable agreement among the results quoted by different authors and the experimental values. The situation is still confused for $A_2$ for which theoretical and experimental results have a large spread.

***B* mesons decays**

The study of $B$ meson decays is of great importance since it is essential for the determination of the still unknown value of $V_{ub}$. $B$ mesons cannot be studied directly on the lattice since the $b$-quark mass is beyond the present values of $a^{-1}$. One possibility is to evaluate the FF in the charm re-gion and to extrapolate the results to the bottom mass. A semi-quantitative study has been performed. We extrapolated by using the behaviour predicted by HQET up to order $1/M_P$ [4]:

$$\frac{(f^+, V, A_2)}{M_P^{1/2}} = \gamma(1 + \frac{\delta}{M_P}); \; A_1 = \frac{\gamma}{M_P^{1/2}}(1 + \frac{\delta}{M_P}) \; .$$

Table 2
FF at $q^2 = 0$ for the $D \to K(K^*)$. The numbers obtained by the ELC and the APE Collaboration are compared to the most recent experimental world average and to previous lattice results. ELC averaged the vector FF obtained from the conserved and the local vector currents[2].

| Ref. | $f^+(0)$ | $V(0)$ | $A_1(0)$ | $A_2(0)$ | $V(0)/A_1(0)$ |
|---|---|---|---|---|---|
| ELC | 0.65(18) | 0.95(34) | 0.63(14) | 0.45(33) | 1.50(28) |
| APE (A) | 0.71(7) | 1.01(15) | 0.67(14) | 0.67(44) | 1.49(33) |
| APE (B) | 0.81(9) | 1.09(27) | 0.72(16) | 0.72(50) | 1.49(33) |
| [5]-[7] | 0.63(8) | 0.86(10) | 0.53(3) | 0.19(21) | 1.6(2) |
| [8]-[9] | 0.90(8)(21) | 1.43(45)(49) | 0.83(14)(28) | 0.59(14)(24) | 1.99(22)(33) |
| [10]- EXP | 0.72(2) | 1.12(16) | 0.59(4) | 0.44(9) | 1.90(25) |

Table 3
ELC (1st row) and Ape (2nd row) results for $B$ FF.

| $f^+(0)$ | $A_1(0)$ | $A_2(0)$ | $V(0)$ | $V/A_1(0)$ |
|---|---|---|---|---|
| 0.33(17) | 0.21(5) | 0.47(28) | 0.40(16) | 1.7(6) |
| 0.35(9) | 0.24(10) | 0.6(1.0) | 0.5(4) | 1.9(1.4) |

In figure 2 we plot the FF as a function of $1/M_P$. The FF for $B$ decays extrapolated at $q^2 = 0$ are given in Table 3.

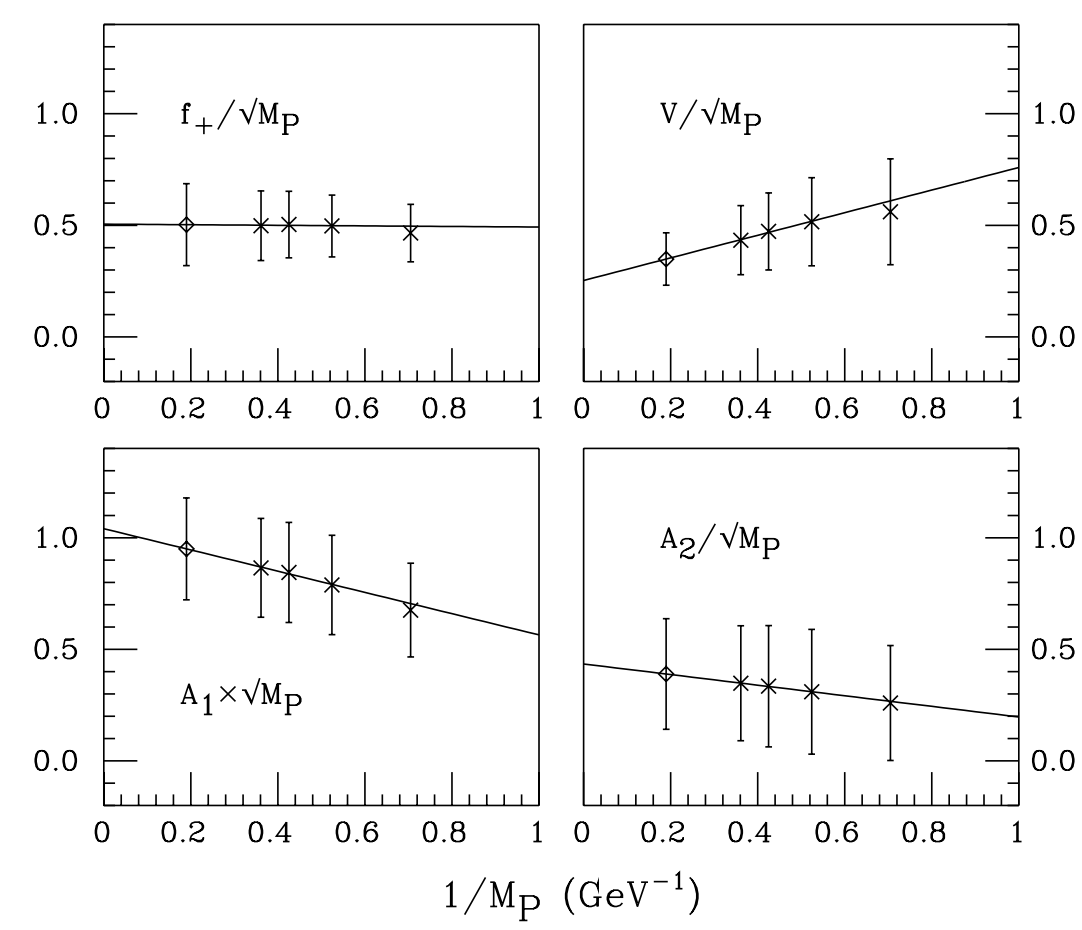

Figure 2. Extrapolation of the FF to the bottom mass (diamond point) using the HQET prediction.